\documentclass[a4paper,11pt]{article}
\usepackage{pos, csquotes}

\title{The Auger Radio Infill SKALA Extension (ARISE): First Measurements of Air Showers}
\ShortTitle{ARISE: First Air-Shower Measurements}

\manuallySeparateAuthors
\author*[a]{Carmen Merx}
\author[b,1]{ for the Pierre Auger Collaboration}

\affiliation[a]{Institute for Astroparticle Physics (IAP), Karlsruhe Institute of Technology (KIT),\\76021 Karlsruhe, Germany}
\affiliation[b]{Observatorio Pierre Auger, Av. San Martín Norte 304, 5613 Malargüe, Argentina}

\emailAdd{carmen.merx@kit.edu}
\emailAdd{spokerspersons@auger.org}

\note{Full author list at \url{https://www.auger.org/archive/authors_2026_06.html}.}

\abstract{The Auger Radio Infill SKALA Extension (ARISE) was installed at the Pierre Auger Observatory in 2025 to detect the radio emission in the band of $50-350\,$MHz of near-vertical cosmic-ray air showers of energies around and above $100\,$PeV. The array comprises 18 SKALA~v2 antennas deployed around one of the surface-detector stations of the $433\,$m array (SD-433), which provides the trigger for ARISE. We present first measurements with ARISE and show the detection of radio pulses associated with extensive air showers. Radio signals are found in coincidence with SD-433 events, and a comparison of the arrival directions reconstructed from the radio and surface detector signals confirms the measurement of air showers with ARISE. These observations provide a first assessment of the detector setup and demonstrate the potential of ARISE for radio detection of near-vertical air showers.}

\FullConference{11th International Workshop on Acoustic and Radio EeV Neutrino Detection Activities (ARENA2026)\\
8-11 June 2026\\
Karlsruhe, Germany\\}

\begin{document}
\maketitle

\section{Introduction}

The Auger Radio Infill SKALA Extension (ARISE) is a new pathfinder radio array installed at the Pierre Auger Observatory~\cite{PierreAuger:2015eyc} in 2025 with commissioning data taking starting in December 2025 and regular data taking starting in April 2026. It consists of six stations, each equipped with three SKALA~v2 antennas \cite{7297231}. The stations are arranged in two concentric circles of approximately $100\,$m and $200\,$m diameter respectively around one of the surface detector (SD) stations, which provides the trigger for data acquisition. Each station features a local data acquisition (DAQ) based on a TAXI board \cite{TAXI}. ARISE is located in the denser area of the Pierre Auger Observatory, designed for the detection of air showers in the transition region between Galactic and extragalactic cosmic rays around $100\,$PeV \cite{Verpoest:2025Oh}.  Further details on ARISE can be found in \cite{SchroederARENA2026}.

\section{Detection of Air Showers with ARISE}

Air-shower events are reconstructed separately with ARISE and with the  SD array with $433\,$m spacing (SD-433) \cite{BrichettoOrquera:202340} of the Pierre Auger Observatory. The reconstructed events are then matched based on their timestamp. The identification of air showers from the radio signal measured by ARISE is performed using the Offline analysis framework~\cite{PierreAuger:2011btp} provided by the Pierre Auger Collaboration.

\subsection{Identification of radio events}

The analysis begins with the identification of air-shower candidate events in ARISE. Radio signals are recorded whenever the surface detector at the center of the array simultaneously triggers all six ARISE stations. The timestamps of this candidate set are subsequently compared to a set of SD events which have passed the trigger criteria of the T3-level \cite{PierreAuger:2005xbq}. Only ARISE events within $5\,\mu\mathrm{s}$ of the T3 level SD events are kept for further analysis.

\begin{figure}[htbp]
\begin{center}
\includegraphics[width=0.9\linewidth]{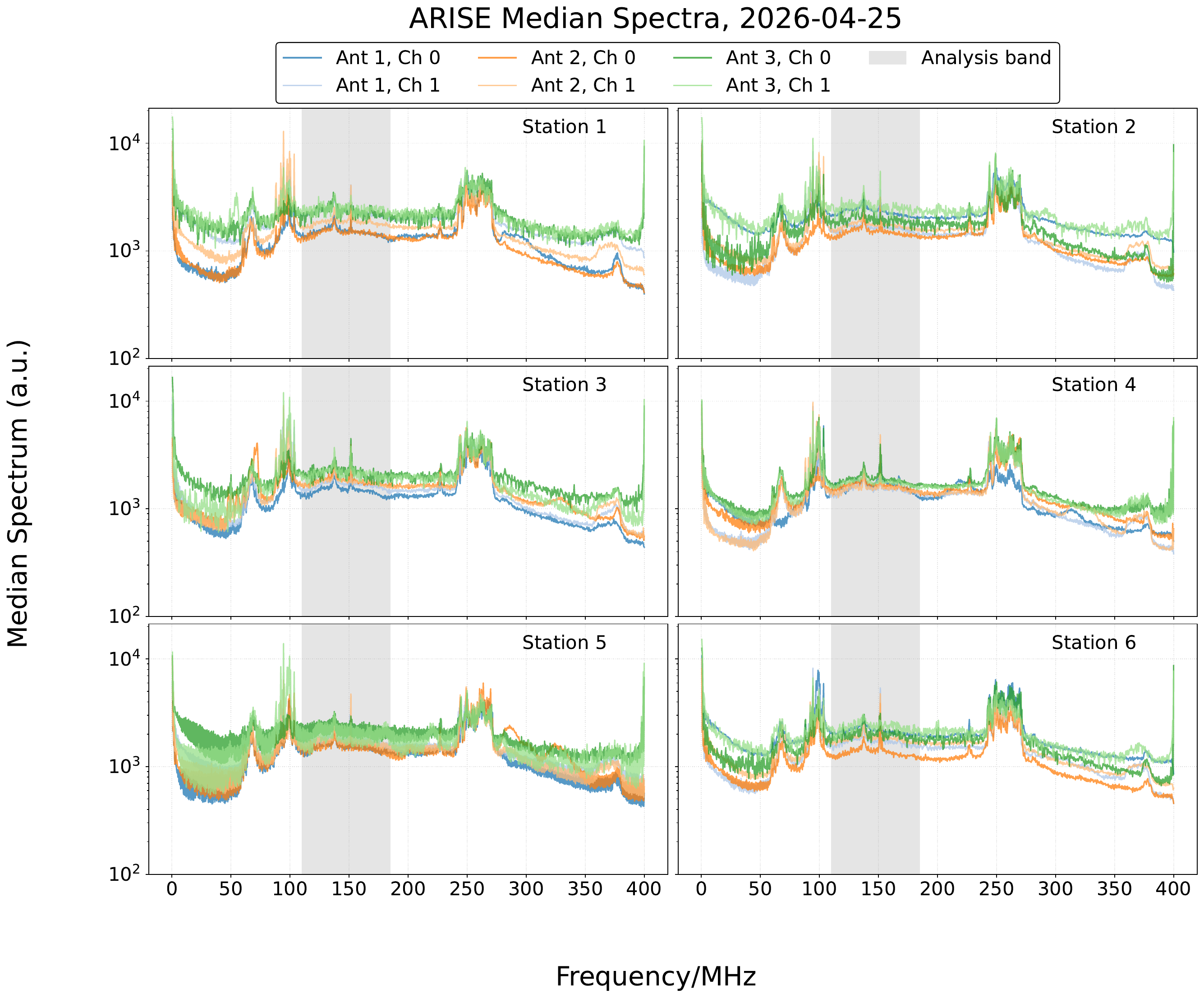}
\caption{ARISE median spectra per station calculated from periodically triggered data.}
\label{spectra}
\end{center}
\end{figure}

After basic preprocessing to remove data-acquisition artifacts, the radio traces are band-limited to 110–185 MHz, a range chosen for its low anthropogenic noise and favorable signal-to-noise ratio relative to the Galactic background (Fig.~\ref{spectra}). Residual narrow-band interferences within this band are subsequently suppressed.

The electric field at each antenna is reconstructed from the recorded voltage traces by unfolding the antenna response in the frequency domain, using the vector effective length that relates the incident field to the induced voltage as a function of frequency and arrival direction. The subsequent analysis is carried out in the two polarization channels, approximately aligned with the east–west and north–south directions. For each channel, we compute the Hilbert envelope of the trace and take the signal time from the channel with the larger peak. Together with the antenna positions, these signal times enter a plane-wave fit that yields an estimate of the air-shower arrival direction.

\subsection{Event Matching with SD-433}

After identifying air-shower candidate events in ARISE, we reconstruct air-shower events for the SD-433 array in the same time interval \cite{BrichettoOrquera:202340}. For each radio event, we search for a coincident SD event within $5\,\mu\mathrm{s}$, after correcting for a system-wide timing offset. In addition, we compare the arrival directions reconstructed by ARISE and the SD-433. If the opening angle between them is below 5°, the radio signals are assumed to originate from the same air shower and the event is used for subsequent data analysis. For this first analysis, we use the shower parameters like timestamp, arrival direction and energy from the SD-433 reconstruction.

\section{First measured events}
\begin{figure}[!htbp]
\begin{center}
\includegraphics[width=\linewidth]{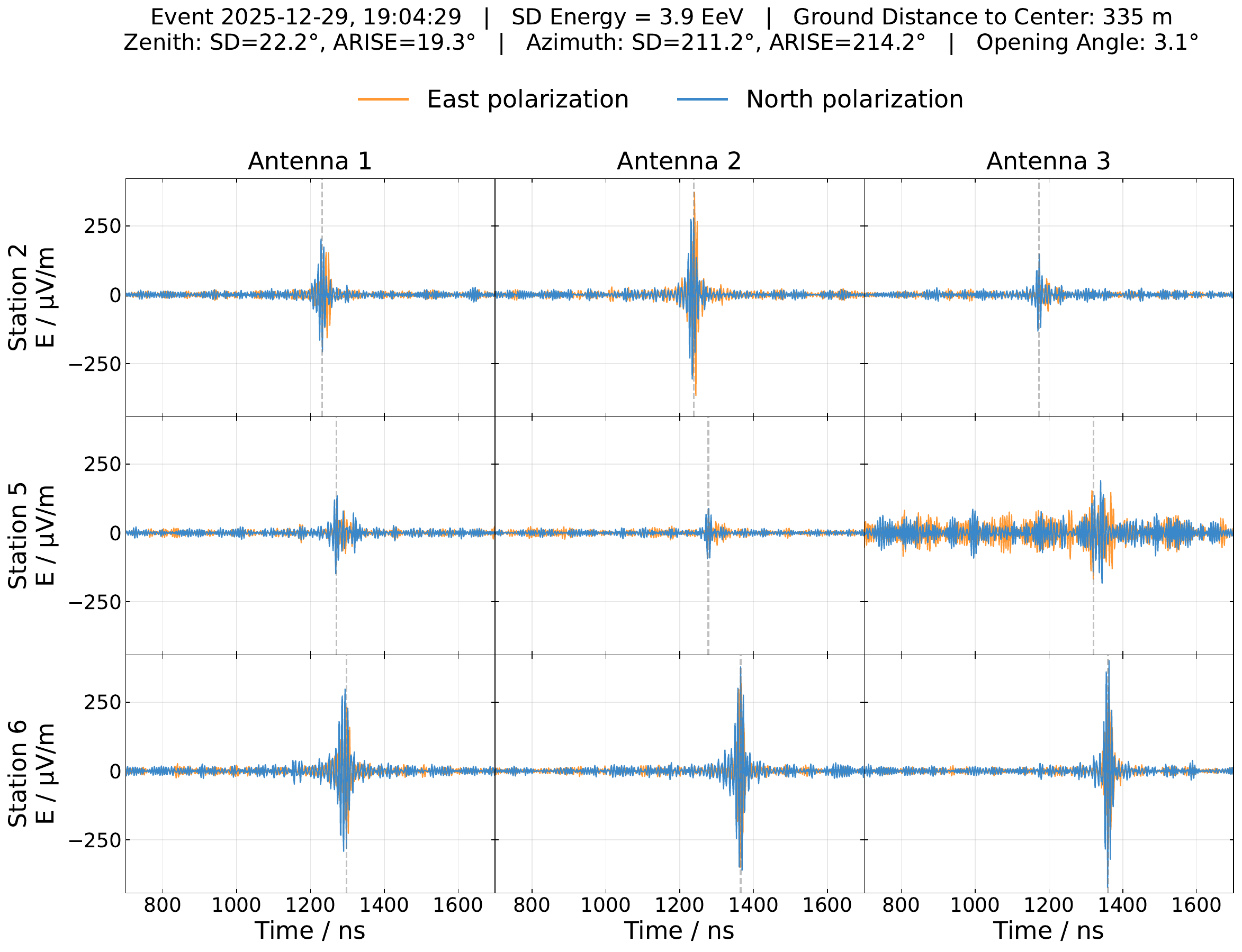}
\caption{Electric-field traces after deconvolution of the antenna response, with a clear signal in all nine antennas operational at the time of the event.}
\label{Waveforms1}
\end{center}
\end{figure}

\begin{figure}[!htbp]
\begin{center}
\includegraphics[width=\linewidth]{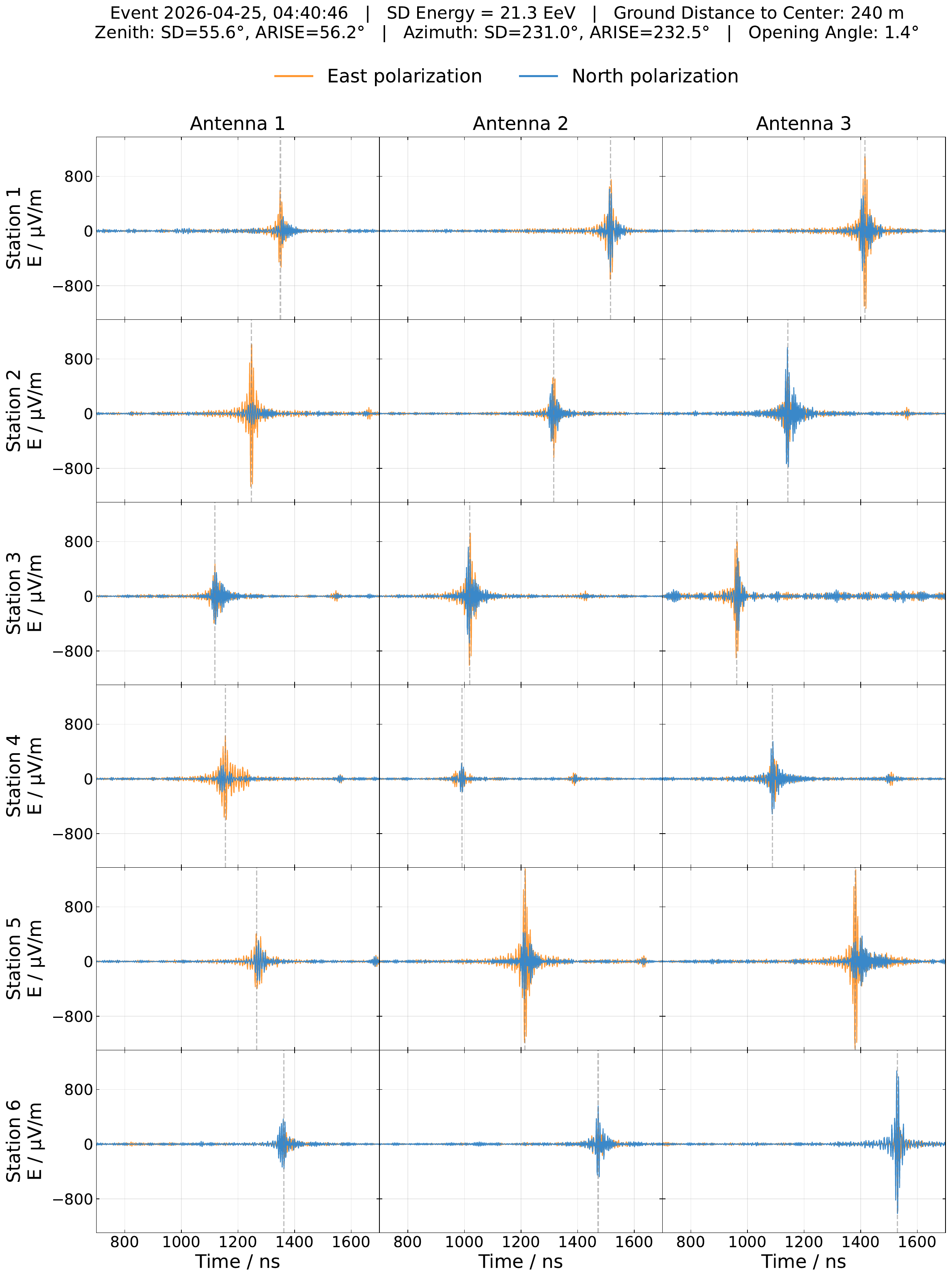}
\caption{Electric-field traces after deconvolution of the antenna response, with a clear signal in all 18 antennas operational at the time of the event.}
\label{Waveforms2}
\end{center}
\end{figure}

We present a first analysis of air-shower events measured with ARISE. The example events shown below were selected from the reconstruction that combines all available stations, whereas the statistical results are based on a per-station approach, in which each station is reconstructed individually. The per-station approach yields more detected events due to better direction reconstruction, which we attribute to the inter-station timing calibration that is still being refined.

We analyzed data recorded between 1 December 2025 and 9 May 2026, selecting events above $100\,$PeV with a shower core within $400\,$m of ground distance of the array center. This yields 181 events using the criteria explained above.

\subsection{Example events}

The two highest-energy coincidences identified in the dataset are shown here in more detail. The first event, shown in Fig.~\ref{Waveforms1}, was recorded on 29 December 2025. It was reconstructed by SD-433 with an energy of $E = 3.9$\,EeV, a zenith angle of $\theta = 22.2^\circ$, and a core located 335\,m from the ARISE array center. At that time, data acquisition was less stable and only three of the six stations were operational. Nevertheless, a clear signal is visible in the radio traces of all nine antennas of the three operational stations. The opening angle between the arrival directions reconstructed by ARISE and SD-433 is $3.1^\circ$.

The second event, shown in Fig.\,\ref{Waveforms2}, was recorded on 25 April 2026. It was reconstructed by the SD-433 array with an energy of $E = 21.3$\,EeV, a zenith angle of $\theta = 55.6^\circ$, and a core distance of 240\,m. Note that the SD-433 reconstruction is optimized for lower-energy events and overestimates the energy for this event, whose energy is approximately $E = 8$\,EeV in the SD-750 and SD-1500 reconstructions. In this case all six stations were operational, yielding a clear signal in all 18 antennas. The reconstructed arrival directions of ARISE and SD-433 agree to within $1.4^\circ$. 

The improved data-taking conditions for the second event reflect a hardware upgrade: in April 2026, the ARISE Control and Housekeeping (CHK) boxes were installed at the stations, enabling monitoring of the station power supply and remote control and power cycling of the DAQ electronics. Data acquisition has been substantially more stable since then, with all six stations operating simultaneously \cite{SchroederARENA2026}.

\begin{figure}[!htbp]
\centering
\includegraphics[width=0.48\textwidth]{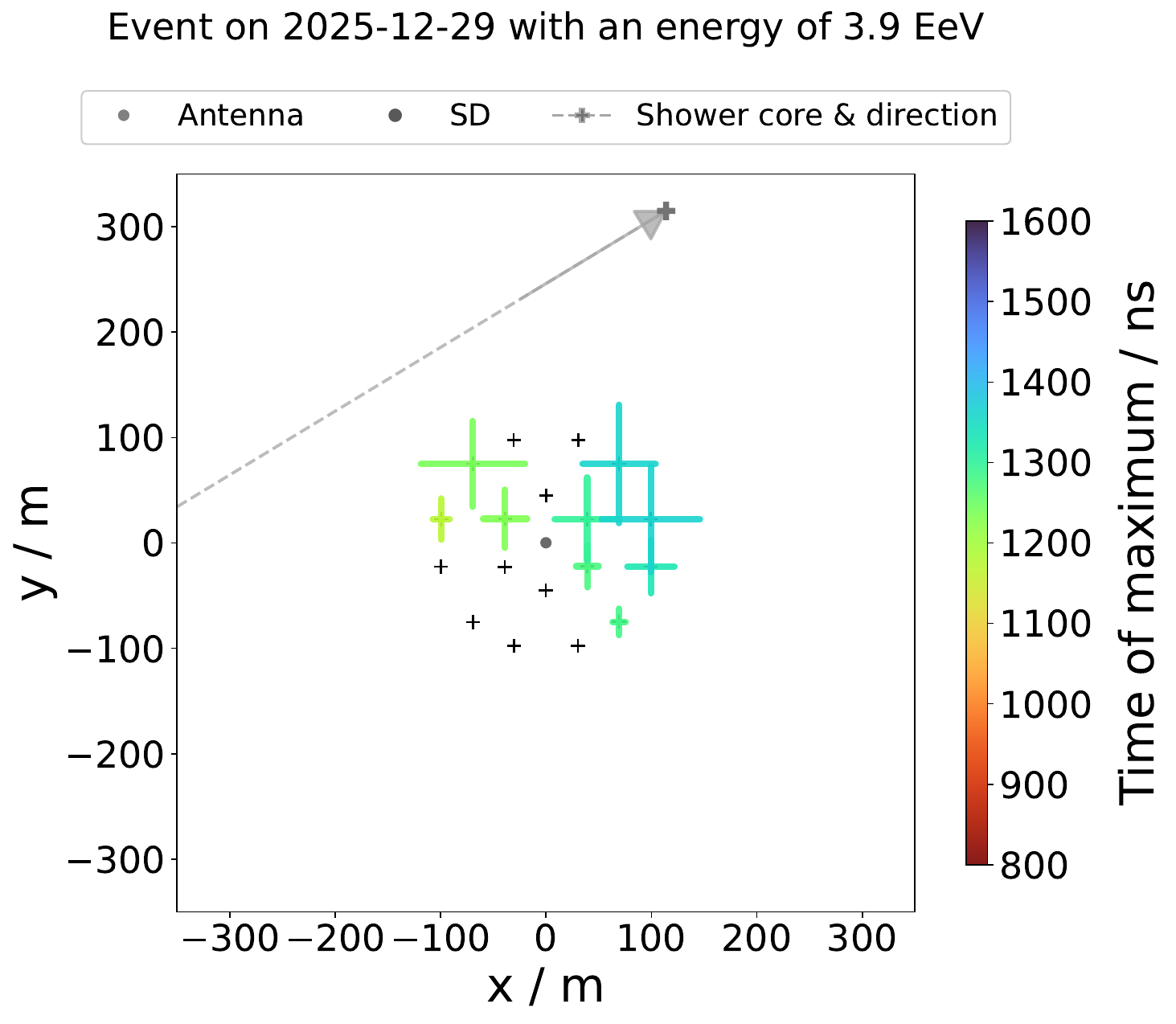}
\hfill
\includegraphics[width=0.48\textwidth]{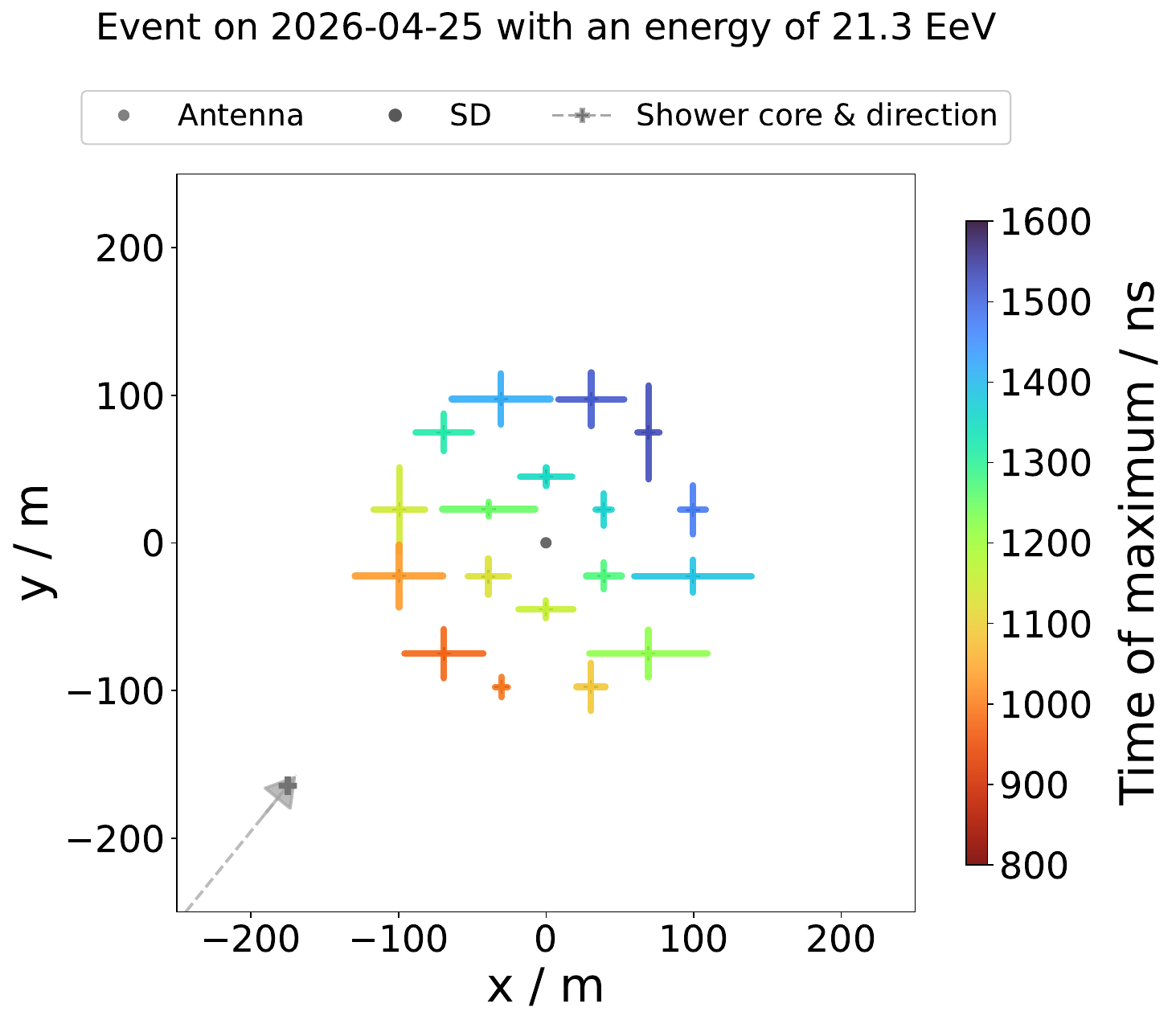}
\caption{These figures show the footprints of the two events. The gray line marks the projected shower axis, with the arrow indicating the propagation direction, and the dark-grey cross marks the reconstructed core position. The active antennas are shown at the center: each is drawn as a cross whose two arms correspond to the north–south and east–west polarizations, with the arm length encoding the signal strength in that channel and the color encoding the arrival time of the signal peak.}
\label{fig:Footprints}
\end{figure}

Fig.~\ref{fig:Footprints} shows the two events. On the left hand side, we can see the three stations that were operational at the time. The arrival-time gradient across the antennas follows the propagation direction of the shower, as expected for an approximately plane wavefront sweeping across the array. In Fig.~\ref{fig:Footprints} on the right, all six stations were operational, giving signals in all 18 antennas.

\begin{figure}[!htbp]
\centering
\begin{minipage}[t]{0.48\textwidth}
\centering
\includegraphics[width=\linewidth]{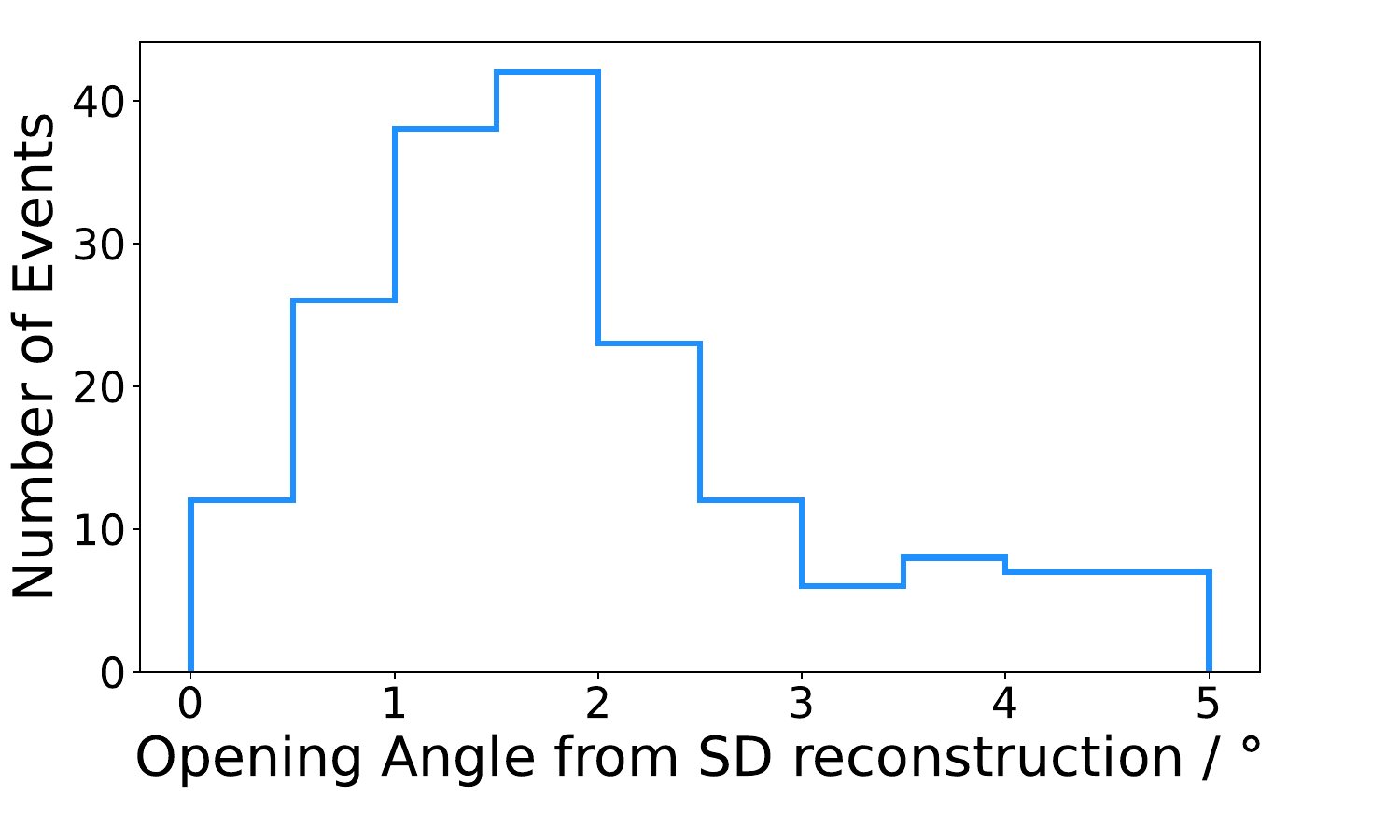}
\caption{Opening angle between arrival directions reconstructed by ARISE with the per-station approach and SD-433.}
\label{fig:openingangle}
\end{minipage}
\hfill
\begin{minipage}[t]{0.48\textwidth}
\centering
\includegraphics[width=\linewidth]{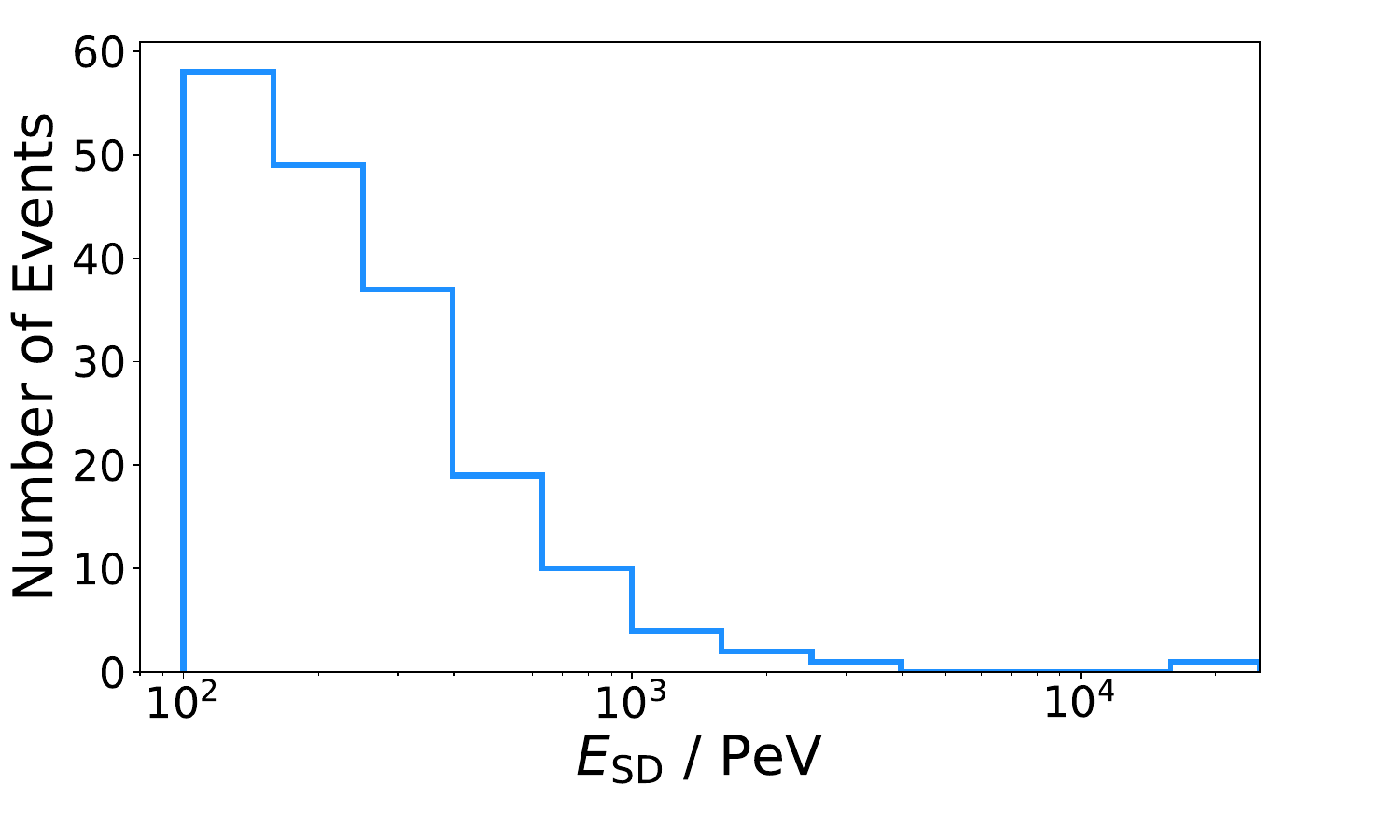}
\caption{Energy distribution of matched events as measured by SD-433.}
\label{fig:energy}
\end{minipage}
\end{figure}

Fig.~\ref{fig:openingangle} shows the distribution of opening angles between the arrival directions reconstructed by ARISE and SD-433. For each event, the ARISE direction is taken from the station whose per-station reconstruction gives the smallest opening angle with respect to the SD-433 direction. The distribution peaks between $1.5^\circ$ and $2^\circ$ which confirms that the radio and surface-detector signals originate from the same air showers. Even with perfect timing, a deviation of $\mathcal{O}(1^\circ)$ may originate from using a plane wavefront in contrast to a more accurate hyperbolic wavefront.

\begin{figure}[!htbp]
    \centering
    \begin{minipage}{0.48\textwidth}
        \centering
        \includegraphics[width=\linewidth]{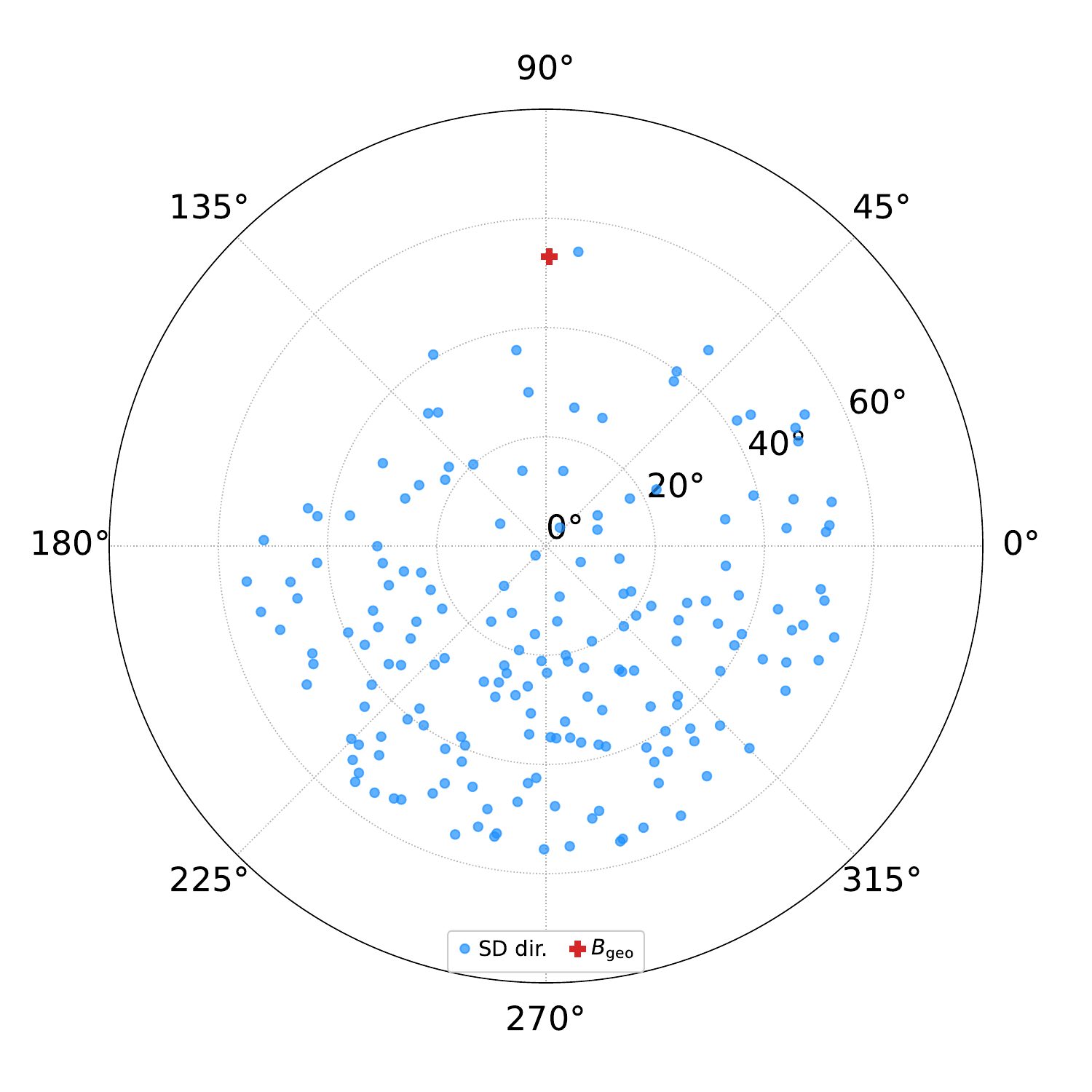}
        \caption{Arrival directions of matched air shower events as reconstructed by SD-433.}
        \label{fig:arrivaldirections}
    \end{minipage}
    \hfill
    \begin{minipage}{0.48\textwidth}
        \centering
        \includegraphics[width=\linewidth]{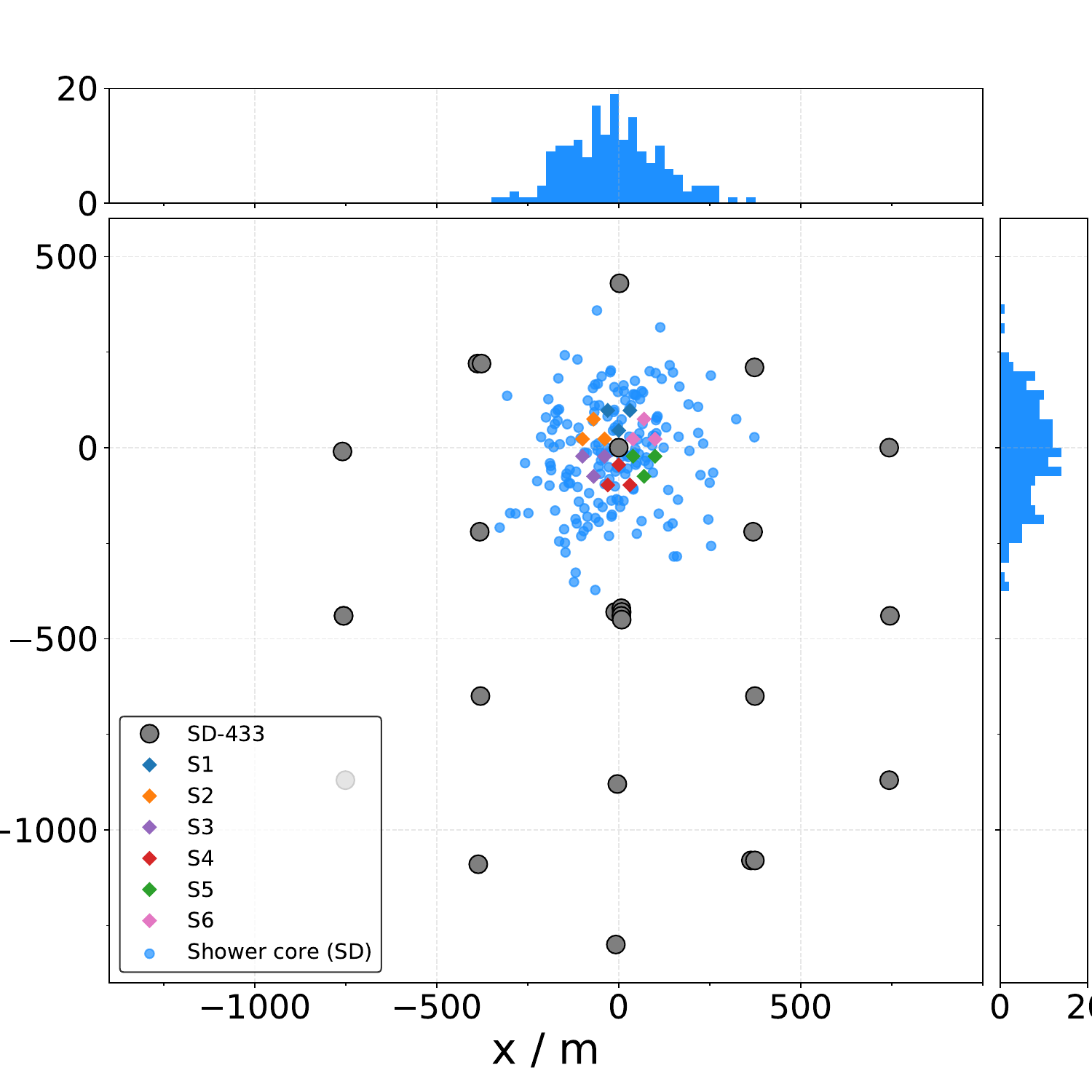}
        \caption{Shower cores with respect to antenna positions of ARISE and SD-433.}
        \label{fig:cores}
    \end{minipage}
\end{figure}

Fig.~\ref{fig:energy} shows the energy distribution of the identified events above the selection threshold of $100\,$PeV. The distribution of reconstructed arrival directions is shown in Fig.~\ref{fig:arrivaldirections}. Events are preferentially detected at large angles to the geomagnetic field, with a deficit for directions close to it. This is expected, since the geomagnetic mechanism, which is dominant for the radio emission of air showers, scales with the sine of the angle between the shower axis and the geomagnetic field. Finally, Fig.~\ref{fig:cores} shows the distribution of reconstructed shower cores relative to the array. In combination, the energy, arrival direction and core distribution clearly indicate that we have detected the radio emission of air showers with ARISE.

\section{Conclusion \& Outlook}

The Auger Radio Infill SKALA Extension is a new antenna array comprising six stations of three SKALA~v2 antennas each at the Pierre Auger Observatory. Analyzing data recorded between December 2025 and May 2026, we identify 181 air-shower events triggered by SD-433 with consistent direction reconstruction, demonstrating that ARISE is capable of detecting cosmic-ray air showers. 

Next steps include refining the inter-station timing calibration to enable a combined reconstruction using signals of all stations simultaneously. In parallel, we plan to simulate the detected air showers with CORSIKA and CoREAS~\cite{Huege:2013vt} to develop a dedicated energy reconstruction for ARISE, and study the array's detection efficiency as a function of the zenith angle and energy of the primary cosmic ray.

\bibliographystyle{ICRC}
\bibliography{bibliography}

@article{PierreAuger:2015eyc,
    author = "Aab, Alexander and others",
    collaboration = "Pierre Auger",
    title = "{The Pierre Auger Cosmic Ray Observatory}",
    eprint = "1502.01323",
    archivePrefix = "arXiv",
    primaryClass = "astro-ph.IM",
    reportNumber = "FERMILAB-PUB-15-034-AD-AE-CD-TD",
    doi = "10.1016/j.nima.2015.06.058",
    journal = "Nucl. Instrum. Meth. A",
    volume = "798",
    pages = "172--213",
    year = "2015"
}

@INPROCEEDINGS{7297231,
  author={de Lera Acedo, E. and others},
  booktitle={2015 International Conference on Electromagnetics in Advanced Applications (ICEAA)}, 
  title={{Evolution of SKALA (SKALA-2), the log-periodic array antenna for the SKA-low instrument}}, 
  year={2015},
  volume={},
  number={},
  pages={839-843},
  doi={10.1109/ICEAA.2015.7297231}
}

@article{SchroederARENA2026,
    author = "Schr{\"o}der, Frank and others",
    collaboration = "Pierre Auger",
    title = "{The Auger Radio Infill SKALA Extension (ARISE): Science Case and Instrumentation}",
    doi = "10.22323/1.538.0002",
    journal = "PoS",
    volume = "ARENA2026",
    pages = "002",
    year = "2026"
}

@article{PierreAuger:2011btp,
    author = "Abreu, P. and others",
    collaboration = "Pierre Auger",
    title = "{Advanced Functionality for Radio Analysis in the Offline Software Framework of the Pierre Auger Observatory}",
    eprint = "1101.4473",
    archivePrefix = "arXiv",
    primaryClass = "astro-ph.IM",
    reportNumber = "FERMILAB-PUB-11-027-AE-CD-TD",
    doi = "10.1016/j.nima.2011.01.049",
    journal = "Nucl. Instrum. Meth. A",
    volume = "635",
    pages = "92--102",
    year = "2011"
}

@article{Huege:2013vt,
    author = "Huege, T. and Ludwig, M. and James, C. W.",
    editor = "Lahmann, Robert and Eberl, Thomas and Graf, Kay and James, Clancy and Huege, Tim and Karg, Timo and Nahnhauer, Rolf",
    title = "{Simulating radio emission from air showers with CoREAS}",
    eprint = "1301.2132",
    archivePrefix = "arXiv",
    primaryClass = "astro-ph.HE",
    doi = "10.1063/1.4807534",
    journal = "AIP Conf. Proc.",
    volume = "1535",
    number = "1",
    pages = "128",
    year = "2013"
}

@article{PierreAuger:2005xbq,
    author = "Allard, D. and others",
    collaboration = "Pierre Auger",
    title = "{The trigger system of the Pierre Auger Surface Detector: operation, efficiency and stablility}",
    doi = "10.48550/arXiv.astro-ph/0510320",
    journal = "PoS",
    volume = "ICRC2005",
    pages = "287",
    year = "2005"
}

@article{TAXI,
    author = "Karg, T. and Haungs, A. and Kleifges, M. and Nahnhauer, R. and Sulanke, K. -H.",
    title = "{Introducing TAXI: a Transportable Array for eXtremely large area Instrumentation studies}",
    booktitle = "{6th International Workshop on Acoustic and Radio EeV Neutrino Detection Activities (ARENA 2014) Annapolis, MD, June 9-12, 2014}",
    journal = "{6th International Workshop on Acoustic and Radio EeV Neutrino Detection Activities (ARENA 2014) Annapolis, MD, June 9-12, 2014}",
    year           = "2014",
    eprint         = "1410.4685",
    archivePrefix  = "arXiv",
    primaryClass   = "astro-ph.IM",
    SLACcitation   = "%%CITATION = ARXIV:1410.4685;%%"
    }

@article{BrichettoOrquera:202340,
  author = "Brichetto Orquera, Gabriel  and  others",
  collaboration = "Pierre Auger",
  title = "{The second knee in the cosmic ray spectrum observed with the surface detector of the Pierre Auger Observatory}",
  doi = "10.22323/1.444.0398",
  journal = "PoS",
  year = 2023,
  volume = "ICRC2023",
  pages = "398"
}

@article{Verpoest:2025Oh,
  author = "Verpoest, Stef  and  Schroeder, Frank  and  Novikov, Alexander  and  Coleman, Alan  and  Flaggs, Benjamin  and  Weindl, Andreas  and  Venugopal, Megha  and  Merx, Carmen  and  Haungs, Andreas",
  title = "{Plans for a new array of radio antennas for the detection of air showers at the 433m surface-detector array of the Pierre Auger Observatory}",
  doi = "10.22323/1.484.0122",
  journal = "PoS",
  year = 2025,
  volume = "UHECR2024",
  pages = "122"
}

\section*{Acknowledgments}
We thank Eloy de Lera Acedo and Quentin Gueuning for their support regarding the SKALA~v2 antennas. 
This project benefited from funding provided by the European Research Council (ERC) and the U.S.~National Science Foundation (NSF). 
Further acknowledgments are included with the full author list at: \url{https://www.auger.org/archive/authors_2026_06.html}

A large language model (Claude, Anthropic) has been used for language editing of this manuscript.

\end{document}